\documentclass[twocolumn,aip,rsi,amsmath,amssymb,reprint,]{revtex4-1}

\usepackage{siunitx}
\usepackage[english]{babel}
\usepackage{graphicx}
\usepackage{mathtools}
\usepackage{subcaption}
\usepackage{nicefrac}
\usepackage{amsmath}
\usepackage{epstopdf}
\usepackage{float}
\usepackage[toc,page]{appendix}
\usepackage{times}
\usepackage{upgreek,textcomp}
\usepackage[labelsep=period]{caption}

\draft 

\draft 

\begin{document}


\title{An Open-Source High-Frequency Lock-in Amplifier} 



\author{G. A. Stimpson}
\email[]{g.stimpson.1@warwick.ac.uk}
\affiliation{Department of Physics, University of Warwick, Coventry,  CV4 7AL, United Kingdom.}
\affiliation{Diamond Science and Technology Centre for Doctoral Training,
University of Warwick,
Gibbet Hill Road,
Coventry,
CV4 7AL,
United Kingdom.}
\author{M. S. Skilbeck}
\affiliation{Department of Physics, University of Warwick, Coventry,  CV4 7AL, United Kingdom.}
\author{R. L. Patel}
\affiliation{Department of Physics, University of Warwick, Coventry,  CV4 7AL, United Kingdom.}
\affiliation{Diamond Science and Technology Centre for Doctoral Training,
University of Warwick,
Gibbet Hill Road,
Coventry,
CV4 7AL,
United Kingdom.}

\author{B. L. Green}
\affiliation{Department of Physics, University of Warwick, Coventry,  CV4 7AL, United Kingdom.}
\author{G. W. Morley}
\email[]{gavin.morley@warwick.ac.uk}
\affiliation{Department of Physics, University of Warwick, Coventry,  CV4 7AL, United Kingdom.}
\affiliation{Diamond Science and Technology Centre for Doctoral Training,
University of Warwick,
Gibbet Hill Road,
Coventry,
CV4 7AL,
United Kingdom.}


\date{\today}

\begin{abstract}
We present characterization of a lock-in amplifier based on a field programmable gate array capable of demodulation at up to $\SI{50}{MHz}$. The system exhibits  $\SI[separate-uncertainty = true]{90}{nV/\surd{Hz}}$ of input noise at an optimum demodulation frequency of $\SI{500}{kHz}$. The passband has a full-width half-maximum of $\SI{2.6}{kHz}$ for modulation frequencies above $\SI{100}{kHz}$. Our code is open source and operates on a commercially available platform.
\end{abstract}


\maketitle 

\section{Introduction}

The lock-in amplifier (LIA) is an invaluable tool in scientific instrumentation\cite{iet1}, allowing the extraction of weak \(\)signals from noisy backgrounds, even where the amplitude of the noise is much greater than the signal \cite{Caplan}. First described in 1941, early LIAs were based on heaters, thermocouples and transformers \cite{Michels}. More modern LIAs have been fully digital \cite{Wang, Probst, hofmann2012, Sonnaillon2005} or implemented on field programmable gate array (FPGA) devices \cite{Bobadilla, giaconia2017, Castillo2005, ieee1, ieee2, divakar2018, chighine2015}, which are able to exceed the performance of their analog counterparts\cite{Carminati2016}. However, the cost of modern LIAs may prove prohibitive, particularly in cases where large numbers of input channels are required. By employing Red Pitaya hardware, we are able to define an open-source LIA solution which is comparable with considerably more costly approaches.
\paragraph*{}

Characterized by wide dynamic range and the ability to extract signal from noisy environments\cite{Meade}, LIAs are phase sensitive detectors\cite{humayun2018} due to their operating principles. A reference signal in the form of a sinusoidal wave is generated either internally by the LIA itself, along with a cosinusoidal wave, or externally by some other source which can also be manipulated to form a cosinusoidal reference. This reference is multiplied by the input signal\cite{scofield1994} which carries the desired data modulated at the reference frequency\cite{horowitz1989}. A low pass filter is then used before outputting the signal. In this way the LIA amplifies and outputs the component of the input signal which is at the reference frequency, attenuating noise at other frequencies\cite{Theocharous:08}. Components out of phase with the sine reference will also be attenuated, due to orthogonality of the sine and cosine functions of equal frequency\cite{devore2016}, hence the LIA is considered phase sensitive \cite{SRSan}. Components in phase with the sine function will result in a non-zero value in the X output, whilst those in phase with the cosine function will produce a non-zero value in the Y output. X and Y can be combined by $R=\sqrt{X^{2} + Y^{2}}$ to provide a magnitude value\cite{ziwp}, while phase $\phi = \arctan{\frac{Y}{X}}$.\\
\paragraph*{}
FPGAs offer easily implementable micro-circuitry design via the use of hardware description language (HDL) such as Verilog or VHDL. A broad range of applications can be realized using these devices \cite{Qian2018} including LIAs without recourse to detailed knowledge of micro-electronics\cite{li2018}.

\paragraph*{}
The Red Pitaya STEMlab 125-14 is a single board computer (SBC) with an integrated FPGA in the form of a Xilinx Zynq 7010 SOC \cite{rpweb}, allowing for the implementation of reprogrammable micro architecture which would otherwise necessitate dedicated hardware. The reprogrammability \cite{Restelli2005} of this FPGA makes it applicable for the LIA functionality described here and allows for further expansion and modification by end users. Two DC coupled analog inputs are available in the form of user selectable $\SI{\pm 1}{V}$ or $\SI{\pm 20}{V}$, $\SI{125}{MS/s}$ analog-to-digital converters (ADCs - Linear Technologies LTC2145CUP-14\cite{adc}) with $\SI{1}{M\Omega}$/$\SI{10}{pF}$ input impedance/capacitance. 

\paragraph*{}
In this article, we detail an LIA which we have implemented on the STEMlab's FPGA chip, along with open source\cite{opensource} operational and data transfer software developed specifically for this application. We demonstrate that this device shares many capabilities with more expensive alternatives such as a sweepable internal signal generator, single or dual input/output modes, wave form control and the ability to increase the number of available inputs and outputs by interfacing across multiple STEMlab units.
\paragraph*{}
Comparison is made with the Zurich Instruments HF2LI LIA, which is specified for operation up to 50 MHz demodulation frequency \cite{hf2lium}. Whilst an extensive software application is provided with the HF2LI, the open source nature and readily available software and hardware of the LIA presented here allow for an attractive option where cost is a consideration. 
Research into low-cost FPGA based LIAs has produced a number of alternatives\cite{li2018}, including high resolution designs operating at up to $\SI{6}{MHz}$ demodulation \cite{Gervasoni2014, Gervasoni2016}, and simulations have been presented for a high frequency LIA based on the Red Pitaya STEMlab\cite{Arnaldi2017}. Typically, FPGA LIAs have been developed with specific experimental objectives \cite{divakar2018, chighine2015, sukekawa2017}. The STEMlab is the basis for a range of related measurement instrumentation from PyRPL, including an LIA\cite{pyrplguiman, pyrplregs, pyrpldacs}. A low cost FPGA-based LIA has also been developed which operates at low demodulation frequency \cite{delacor}, and FPGA-based LIAs have been compared with analog devices in terms of signal accuracy\cite{vandenbussche}. However, we believe that this article is the first to characterize a high frequency, open source LIA. The open source code may lead to a range of future uses in research, education and industry.  
\section{Methods}
\begin{figure*}[ht]
    \centering
    \includegraphics[width=0.8\linewidth]{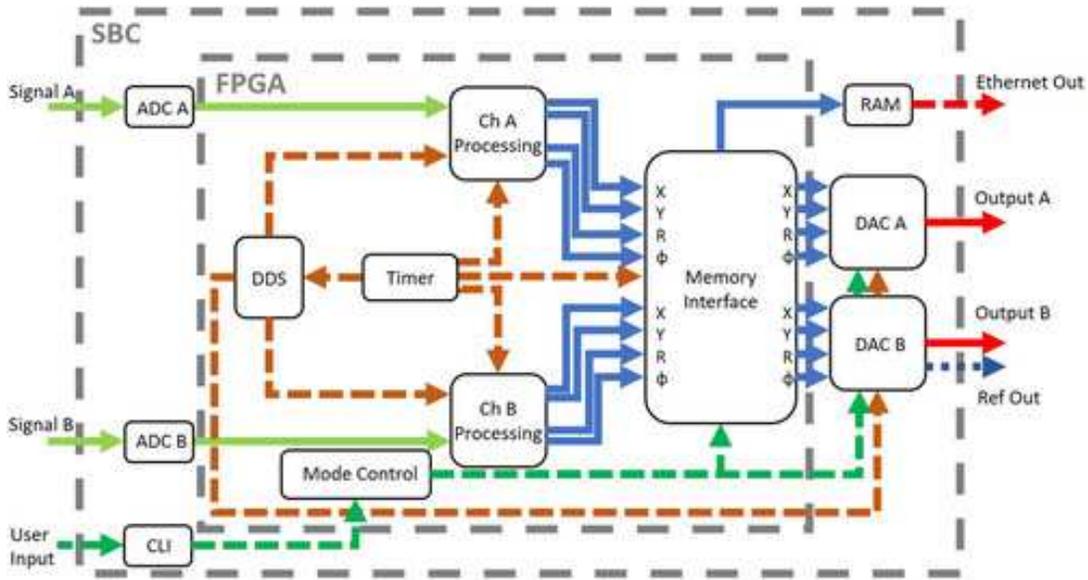}
    \caption{Simplified schematic of circuitry layout on Red Pitaya lock-in amplifier FPGA. Signals (solid green arrows) are received at channel processing, where they are multiplied by a reference signal generated in the direct digital synthesis (DDS) block, which takes time data (dashed brown arrows) from the timer and distributes it as a reference. After the multiplication, data are passed through a single pole infinite impulse response (IIR) filter, currently limited to single order. Filtered data (solid blue arrows) are then passed to the memory allocation block and subsequently to the digital to analogue converters (DACs) and are converted to analog data (solid red arrows) and passed out to the user, or saved to the SBC RAM where they can be accessed via Ethernet (dashed red arrow). The reference signal is extracted from the DACs in analog form (dashed blue arrow). Operating parameters and modes are set by the user (dashed green arrows), via the command line interface (CLI) which communicates with the mode control block on the FPGA.}
    \label{fig:my_label}
\end{figure*}

\subsection{Hardware \& Software}
The FPGA system design and implementation was performed using software provided by the manufacturer of the FPGA, Xilinx. A simplified version of the design circuitry block diagram can be seen in fig. 1.  Signals received by the ADCs are passed to processing blocks where they are multiplied by the reference signal (which is generated internally by direct digital synthesis (DDS)) and filtered using a single pole infinite impulse response (IIR) filter. The resulting output is written to the SBC's random access memory (RAM) via the FPGA's memory interface block. These data are written to a ramdisk file also contained within the STEMlab's RAM, alleviating high read/write workloads which were observed to cause critical failures of the SD card. The data are also passed to the on-board digital-to-analog converters (DACs) along with the reference signal which is used to modulate the desired signal. This reference can be extracted via the DAC output for external use.

\paragraph*{}
Data retrieval from the STEMlab to a host computer can be achieved via the DACs which are provided with SMA connections for output to an oscilloscope. These DACs have 14-bit resolution combined with 125 MS/s data rate. Non-offset, digital amplification of up to two thousand times the raw output is available. However, noise introduced by the DACs makes the output undesirable in cases where small changes in signal intensity are to be detected. Alternatively, data may be transferred to a host PC via file transfer from the STEMlab's RAM. This produces low noise data but can only be performed on the entirety of the LIA's allocated storage space of approximately 65 megabytes (MB). Whilst this process may last for tens of seconds, all data (X, Y, R and $\phi$) for both output channels is received simultaneously. Further limitations include the inability to operate using an external lock-in reference, cross-talk which can occur between the two input channels and no facility for subtraction of channels (e.g. A-B). Due to the programmable nature of the STEMlab, end users are able to implement their own data transfer methods, which may be shown to alleviate some or all of these limitations.

\paragraph*{}
Whilst the FPGA is responsible for signal processing and data transfer, parameter setting takes place on the STEMlab's Linux SBC. This device is packaged along with the FPGA on the STEMlab board, allowing the user to operate the Red Pitaya LIA (RePLIA) or any other custom FPGA application entirely using the STEMlab itself. Parameter setting and system control are performed using C and Python codes written specifically for use with the LIA. These codes may be accessed via SSH or a terminal emulator, but in practice are accessed via a Java-based graphical user interface (GUI) on a client computer which allows the user to largely ignore the STEMlab's Linux element. This open-source GUI allows for the setting of all available RePLIA parameters.

\subsection{Characterization Methodology}
Data were extracted from both LIAs via Ethernet connection. In the case of the RePLIA aboard the STEMlab, this avoided noise from its DACs which were found to increase noise by up to three orders of magnitude. 
\paragraph*{}
Noise for both LIAs during operation was measured at various demodulation frequencies with no input signal present and with a constant amplitude signal produced with an Agilent N5172B EXG vector signal generator. Similarly, noise was measured whilst varying the time constant of the LIAs, with the demodulation frequency fixed at the value determined to be the least noisy by the previous method. As with input noise, these data are presented after a fast Fourier transform (FFT).

Passbands for each LIA at various demodulation frequencies were obtained by applying a constant amplitude signal at the specified demodulation frequency. This signal was then combined with a sweep between a minimum and maximum frequency passing through the demodulation frequency, with a $\SI{10}{s}$ sweep time.

\subsubsection*{Zurich Instruments HF2LI}
The HF2LI provides a built in input noise measurement protocol. The experimental conditions detailed in the HF2LI user notes were replicated\cite{hf2lium} and the input noise measured. Noise was collected whilst inputs were terminated for periods of 1, 10 and 100 seconds. This noise was then subjected to a FFT and the noise level assessed and compared with the manufacturer's stated input noise level of $\SI{5} {nV/\surd{Hz}}$ at $\SI{1}{MHz}$ demodulation for output frequencies greater than $\SI{10}{kHz}$\cite{zispecs}.

\subsubsection*{Red Pitaya Lock-In Amplifier}
Optimal operational parameters were deduced by examining the output of the RePLIA at varying demodulation frequencies and time constants. The optima chosen were those which resulted in the lowest output noise level. This behaviour was examined both with an applied signal and  with terminated inputs. The RePLIA maintains capability over a wide range of demodulation frequencies ($\SI{10}{kHz} - \SI{50}{MHz}$) which were also characterized, but figures are quoted with respect to optimal parameters unless otherwise stated.

\section{Results}
Figure 2 shows input noise spectral density for both LIAs at 1 second, 10 second and 100 second collection periods. Increasing collection time $t$ was expected to produce a $\sqrt{t}$ decrease in noise. Both LIAs remained within close agreement of this prediction after a 100 seconds collection time (fig. 2c). noise levels were obtained by taking an average of base-lined data from each LIA at the relevant collection time. In figure 2a, the data for 1 second of collection time has been zero padded, adding 4 seconds worth of zeros, to reduce the width of the frequency bins. These averages produce figures of $89$, $21$ and $\SI{6}{nV}$ for the RePLIA for collection times of 1, 10 and 100 seconds respectively, corresponding to sensitivities of $89$, $66$ and $\SI{60}{nV/\sqrt{Hz}}$. When the RePLIA was operated using the conditions stipulated for measurement of the HF2LI input noise, the noise levels are $131$, $133$ and $\SI{130}{nV/\sqrt{Hz}}$. This shows that the optimal operating parameters for the HF2LI differ from those appropriate to the RePLIA. It should be noted that the manufacturer specified sampling rate for the HF2LI limits the span of the FFT output to $\SI{120}{Hz}$. Frequencies beyond this limit can be found in the HF2LI's user manual in scetion 8.4 page 665\cite{hf2lium}. A wider output frequency plot can be found in the appendix (figure \ref{fig:rpwide}).
\begin{figure}[ht]
\includegraphics[width=0.8\linewidth]{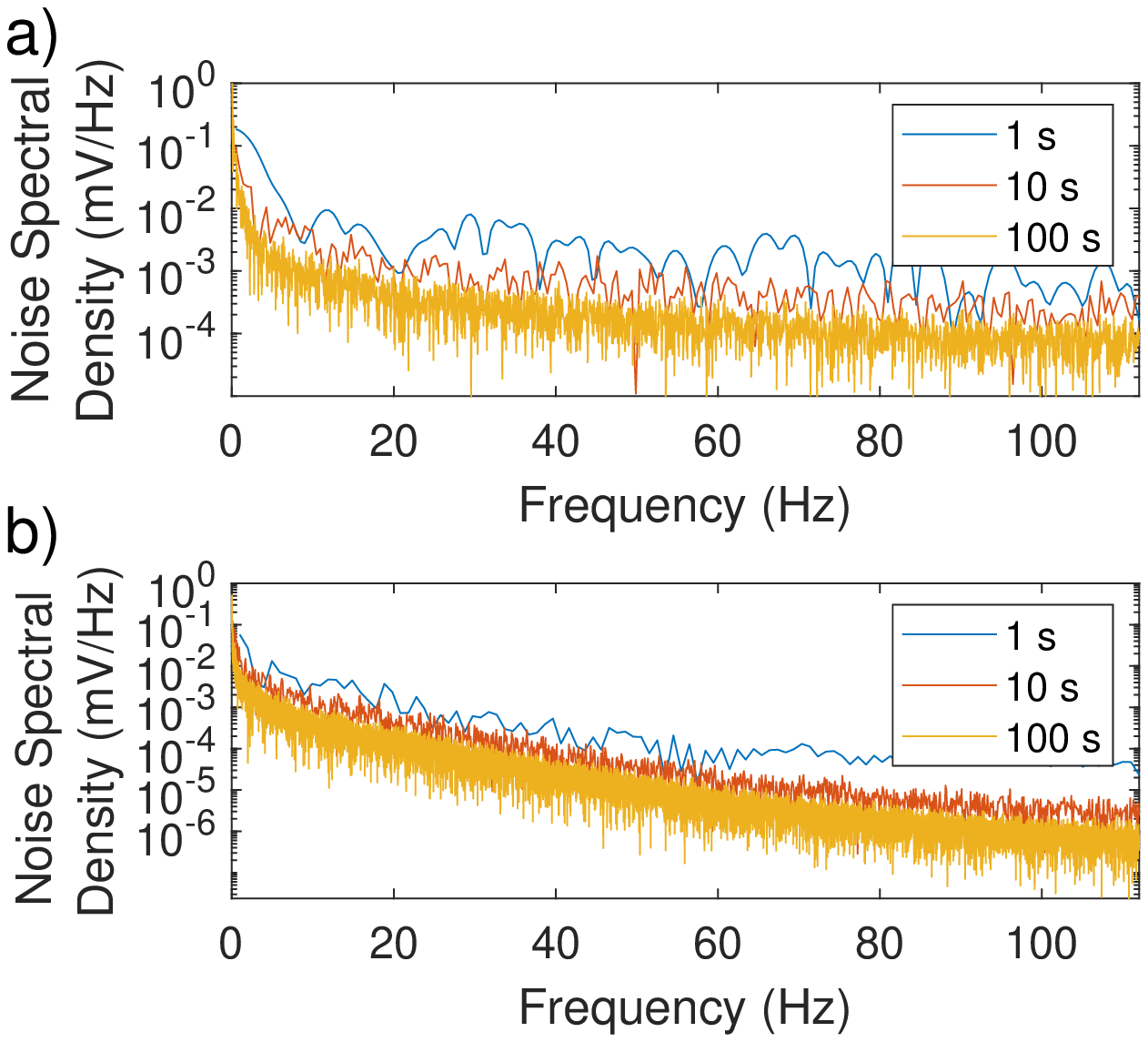}
\includegraphics[width=0.8\linewidth]{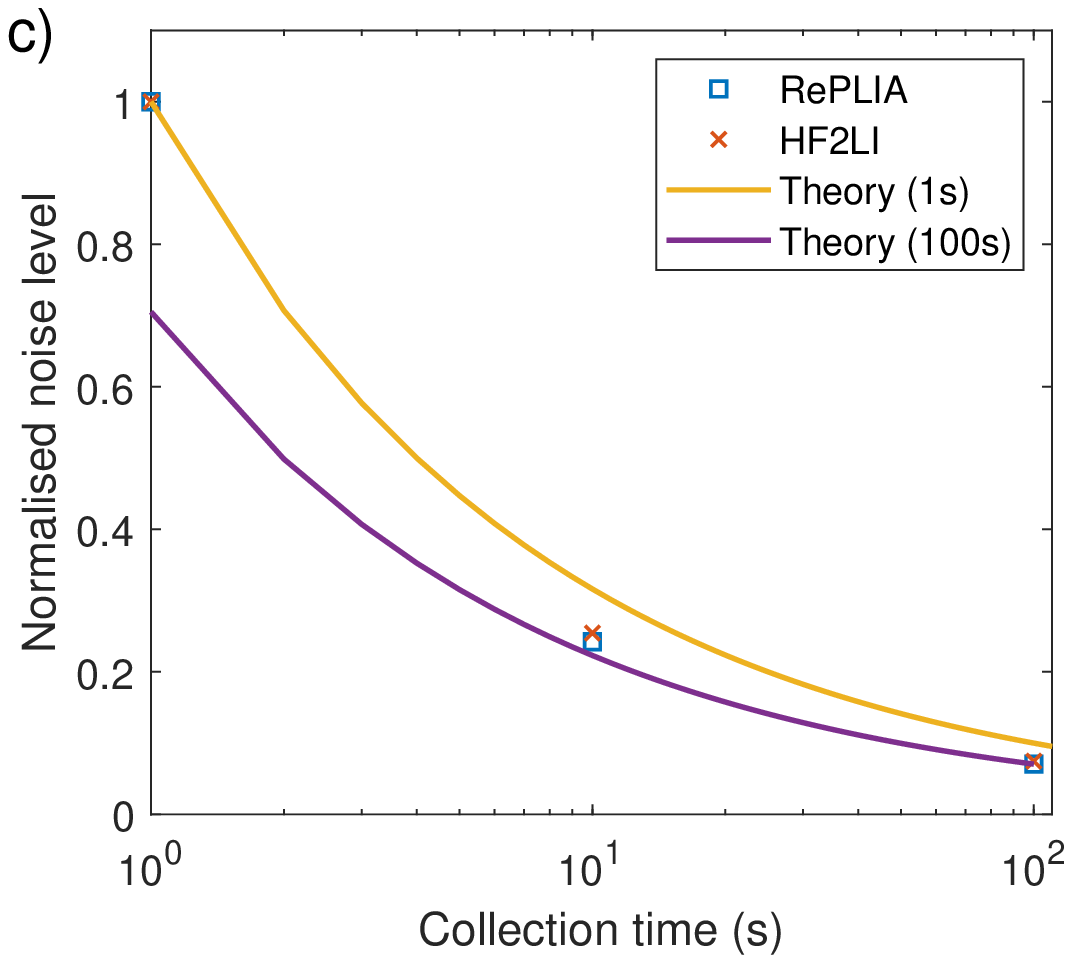}
\caption{Noise spectral density from a) RePLIA at $\SI{500}{kHz}$ demodulation with a $\SI{1}{ms}$ time constant (with $\SI{1}{s}$ data zero padded) and b) HF2LI at $\SI{1}{MHz}$ demodulation and $\SI{700}{\micro s}$ time constant, at 1 second, 10 seconds and 100 seconds collection time (output data is R for both channels). c) noise level decays with increased collection time, close to the predicted $\sqrt{t}$ relationship (top trace) calculated based on noise measured after 1 second of collection. The lower trace indicates theoretical values back calculated from the 100 seconds data point.} \label{fig:combinednoise}
\end{figure}
\paragraph*{}
The noise spectrum for the HF2LI (fig. 2b) shows greater noise below $\SI{80}{Hz}$ and lower noise at higher output frequencies. By contrast, the RePLIA shows a higher level of noise which is largely steady up to 120 Hz (fig. 2a) output frequency.
\paragraph*{}
Figure 3 shows FFTs of the RePLIA output collected for $\SI{100}{s}$ at a range of demodulation frequencies, with time constants optimized for the relevant demodulation period (see appendix). There is a marked difference in noise at demodulation frequencies below $\SI{100}{kHz}$, though above this threshold noise is roughly constant. 

\begin{figure}
\centering
\includegraphics[width=\linewidth]{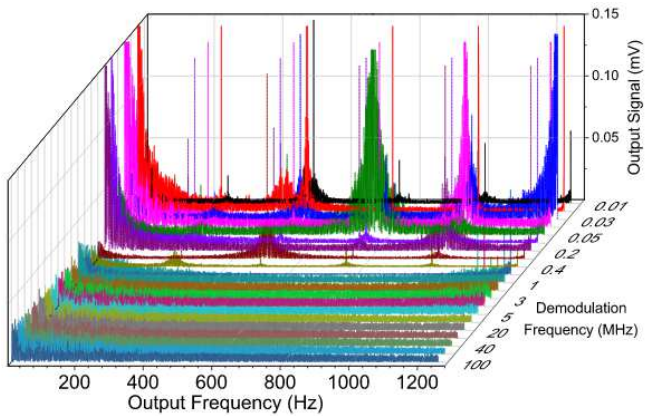} 
\caption{FFTs of terminated input data, extracted via file transfer, at various RePLIA demodulation frequencies with $\SI{100}{s}$ collection time. Two distinct noise regimes seem to exist, with demodulation frequencies above $\SI{100}{kHz}$ producing significantly lower noise. Time constants were varied dependent on demodulation frequency (see appendix). Note that low frequency and high intensity data have been cropped for visual clarity.}
\end{figure}

Output from the RePLIA digital-to-analogue converters (DACs) was also produced (see appendix), showing a constant noise level at frequencies below 100 kHz. Though this noise level is several orders of magnitude greater than that incurred in data extracted via the Ethernet connection, this may be an acceptable trade off in certain cases. This consistency suggests that noise inherent to the DACs is the limiting factor when obtaining data using this method, though this noise can be reduced as described in reference 28. Signal-to-noise as measured at the DACs increases with an increase in the DAC multipliers, plateauing towards saturation of the RePLIA's outputs (see fig. 7 in the appendices).

\paragraph*{}
Figures 4a and 4b show the shape of LIA passbands when modulating at varying frequencies with time constants optimized for each frequency. The HF2LI demonstrates a  narrower passband with less transmission at unwanted frequencies. These passbands are shown to be approximately Lorentzian in line shape with full-width half-maxima of $\SI{2.6}{kHz}$ for demodulation frequencies above $\SI{100}{kHz}$ (see appendix), and they begin to reduce in intensity above $\SI{30}{MHz}$ demodulation with the HF2LI, whereas the output is consistent up to $\SI{50}{MHz}$ demodulation with the RePLIA. Figure 5c shows the normalised output at megahertz demodulation frequencies for both LIAs. The  half power point is indicated ($\frac{V_{max}}{\sqrt{2}}$) (where $V_{max}$ is the peak voltage output), which for the RePLIA is encountered at a demodulation frequency of around 60 MHz, which is close to half of the device's $\SI{125}{MHz}$ sample rate, as can be expected from the Nyquist sampling theorem \cite{Shannon}.  

\paragraph*{}
Linearity of the output relative to the input is demonstrated in figure 5a. Input voltages were varied and the resultant DAC output voltage measured for both the $\SI{\pm 1}{V}$ and $\SI{\pm 20}{V}$ input ranges. Linearity is observed over $\SI{\pm 700}{mV}$ and $\SI{\pm 1100}{mV}$ respectively. Outside these regions the STEMlab's input, having only 14 bits, introduces significant non-linearities, as does improper impedance matching of the RePLIA's inputs.

\begin{figure}
\centering
\includegraphics[width=0.48\linewidth]{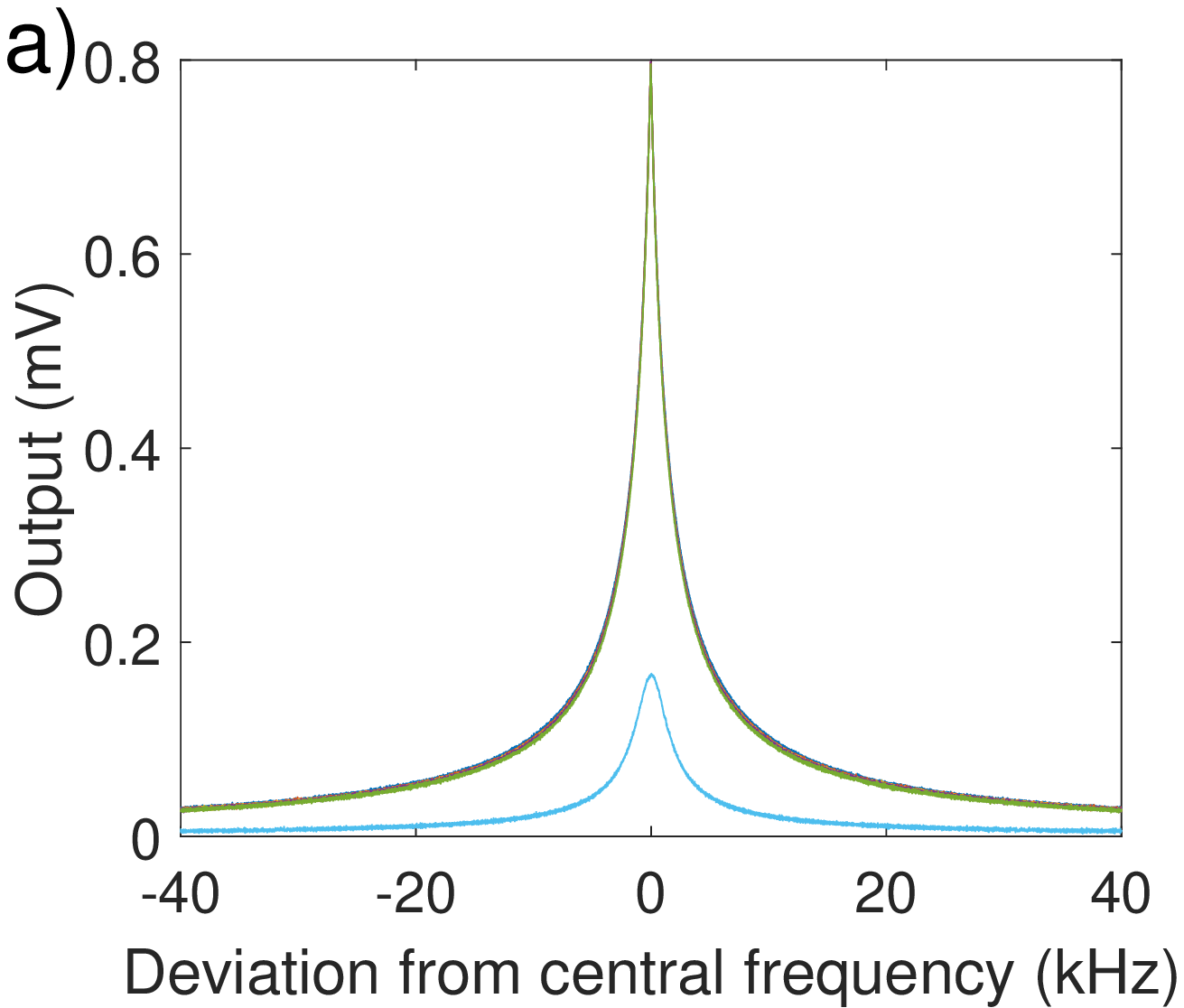}
\includegraphics[width=0.48\linewidth]{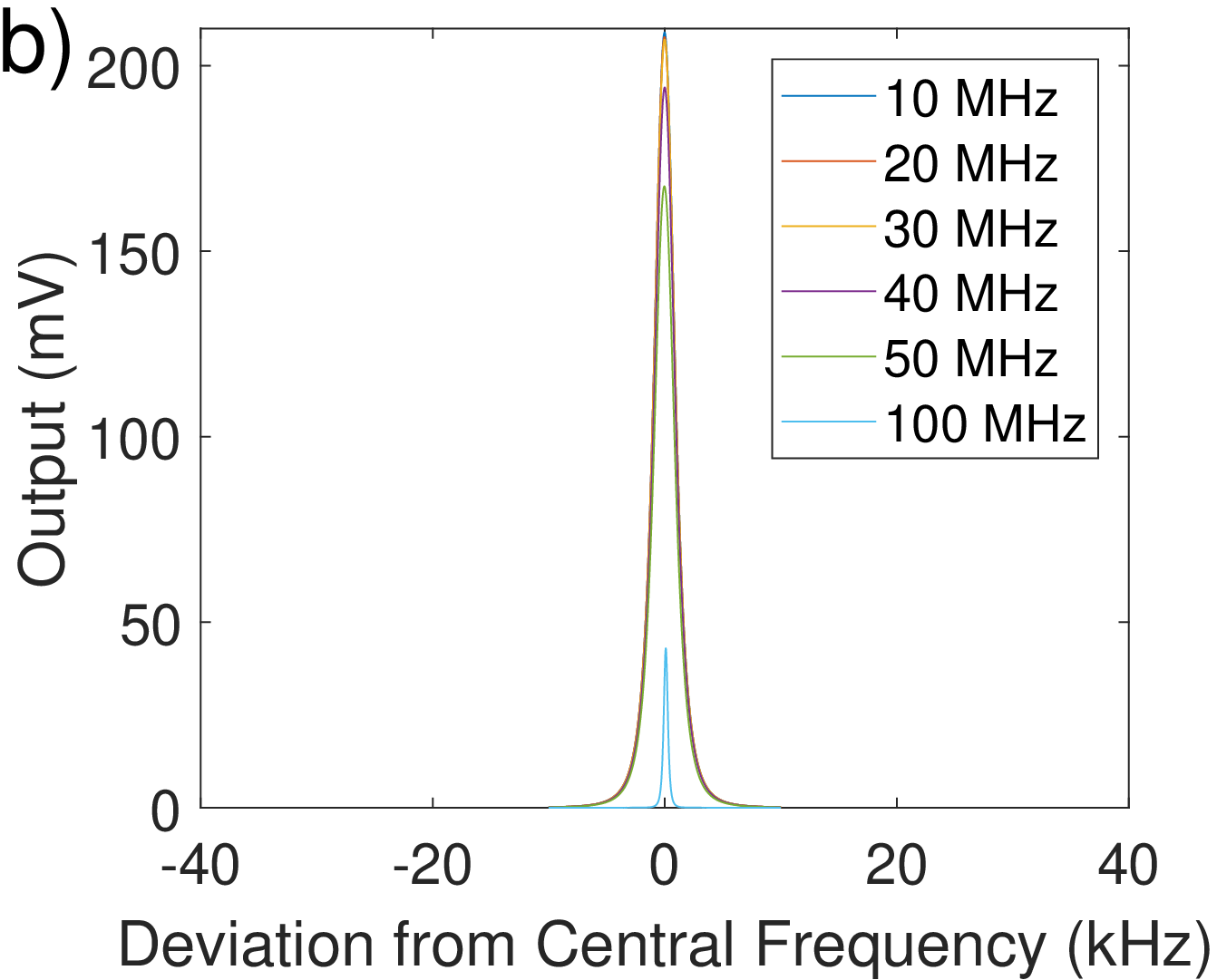}
\includegraphics[width=0.3\textwidth]{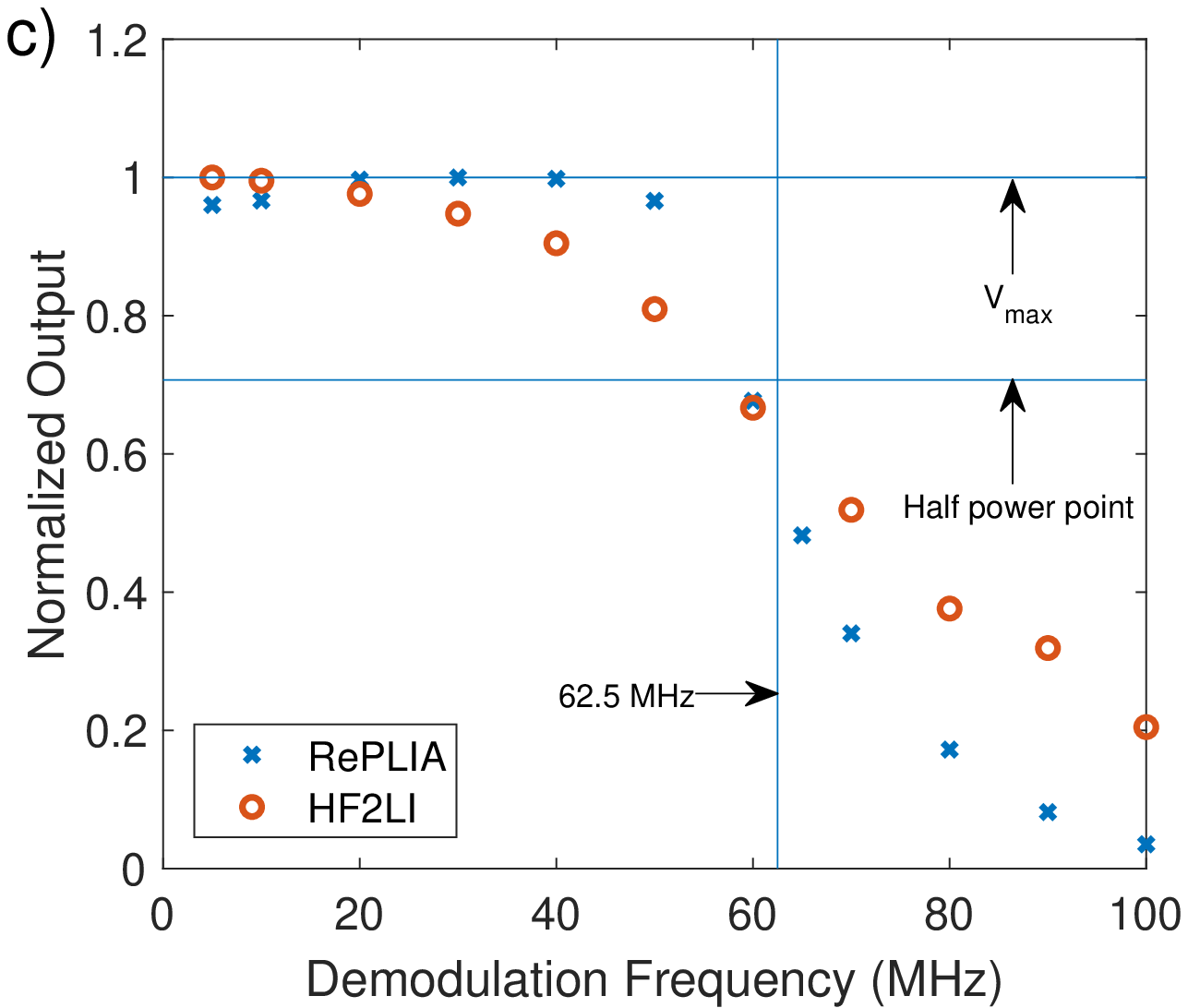}
\caption{Output of frequency sweep for MHz demodulation frequencies for the a) RePLIA and b) HF2LI. c) Maximum output voltage of the RePLIA plotted against demodulation frequency, showing the half power point. Note that the RePLIA has a sampling rate of 125 MHz.}
\end{figure}

\begin{figure}
    \centering
    \includegraphics[width=0.68\linewidth]{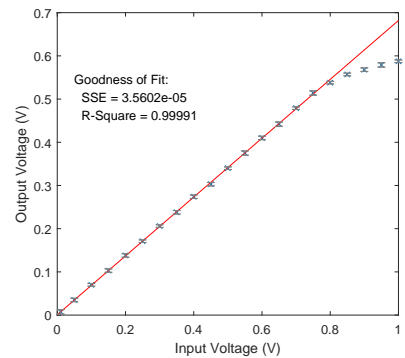}
    \caption{Demodulated RePLIA output against input voltage, operating at $\SI{500}{kHz}$ demodulation for a $\SI{\pm 1}{V}$ input range. Solid red line is a linear fit of the 0 to $\SI{0.7}{V}$ linear region.}
\end{figure}

\section{Conclusions}
The RePLIA performs lock-in amplification at frequencies up to $\SI{50}{MHz}$, with an input noise level of $\SI[separate-uncertainty = true]{90}{nV/\surd{Hz}}$. The respective figures from the Zurich Instruments HF2LI ($\SI{50}{MHz}$ demodulation with $\SI{5}{nV/\surd{Hz}}$ input noise) show that the FPGA LIA detailed in this article is an open source alternative. 
\section*{Supplementary Material}
An online appendix presented alongside this article provides operational parameters and methodology pertaining to both the RePLIA and HF2LI. Also contained within this material are further data regarding measurements of DAC noise at varying demodulation frequency and amplification levels, as well as output line shape and lock-in time constant behaviour.
\section*{Acknowledgements}
We thank Ben Breeze and Matt Dale for assistance and advice. This research was funded by Bruker Biospin and the Engineering \& Physical Sciences Research Council (EPSRC) via grant EP/L015315/1, and G. W. Morley is supported by the Royal Society. The data for this report is open access and is available on the University of Warwick Publication Service \& WRAP at http://wrap.warwick.ac.uk/124332/.
\begin{appendix}
\section{Operational Notes}
The RePLIA was operated in dual channel input mode with the output producing amplitude data. These data were collected at a rate of $\SI{20}{kSamples/sec}$. It is important to note that in setting a sample rate, the actual rate per sample will be $\nicefrac{1}{8}$ of that set, as the set rate includes one sample per output channel per time increment (2 input channels comprising R, X, Y and $\phi$ for each input channel). This sample rate is dependent both on the system being measured and the hardware over which data are transmitted, but can be set by the user to any rate within the sample rate range of the STEMlab, although this sample rate may ultimately effect the accuracy of resultant signals. When obtaining data via the DAC outputs, collection has been performed by a Pico  Technology Picoscope 4424 digital oscilloscope. The Picoscope and DAC outputs were also used when determining time constants for the RePLIA. Data extracted via the RePLIA's Ethernet port were converted to millivolt values by dividing the actual numerical value by $2.1\cdot10^{6}$, a conversion factor which was determined by comparing Ethernet values with corresponding values produced at the DACs with no DAC multiplication applied.
\paragraph*{}
The Zurich Instruments HF2LI was operated according to the instructions laid out in section 3.5 of the HF2 User Manual\cite{hf2lium}, using the parameters in Table I.
\\
\\
\begin{table}
\begin{tabular}{ |p{3cm}|p{2cm}|p{2cm}|}
 \hline
 \multicolumn{3}{|c|}{HF2LI Input Noise Settings} \\
 \hline
 Setting & Parameter & Value  \\  
 \hline\hline
 Signal Input & Range & $\SI{10}{mV}$  \\ 
 \hline
  & AC & On \\ 
 \hline
 & Diff & Off  \\
 \hline
 & $\SI{50}{\Omega}$ & On  \\
 \hline
 Demodulator Low-pass filter& Bandwidth 3dB & $\SI{100}{Hz}$  \\
 \hline
 & Order & 4  \\
 \hline
 Oscillator & Frequency & $\SI{1}{MHz}$ \\
 \hline
 Signal Output & Switch & Off \\[1ex]
 \hline
\end{tabular}
\caption{Experimental parameters used for the measurement of HF2LI input noise, as detailed in manufacturer’s instructions.} 
\end{table}

\paragraph*{}
The time constants detailed in Table II were used for measurements made with the RePLIA. For frequencies in between those listed, the time constant corresponding to the lower frequency was used. Minimum time constants $\tau_{min}$ have been determined by $\tau_{min}=10/f_{mod}$ where $f_{mod}$ is the demodulation frequency. This ensures that for a given time constant, ten modulation periods or more elapse during the time constant. The actual time constants used were chosen based on the lineshape of passbands as seen in figure 5a in the main text. Time constants were made as large as possible without distortion of the Lorentzian lineshape of these passbands. 
\\
\\
\begin{table}
\centering
\begin{tabular}{ |p{2cm}|p{2cm}|p{2cm}|}
 \hline
 \multicolumn{3}{|c|}{Time Constants ($\tau$)} \\
 \hline
  $F_{mod} (\mathrm{MHz})$ & $\tau_{min} (\mathrm{ms})$ & $\tau (\mathrm{ms})$ \\  
 \hline\hline
 0.001 & 10 & 100 \\ 
 \hline
 0.01 & 1 & 100 \\ 
 \hline
 0.1 & 0.1 & 10 \\
 \hline
 0.5 & $2\cdot10^{-2}$ & 1 \\
 \hline
 1 & $1\cdot10^{-2}$ & 1 \\
 \hline
 10 & $1\cdot10^{-3}$ & 1 \\
 \hline
 100 & $1\cdot10^{-4}$ & 0.1\\[1ex] 
 \hline
\end{tabular} 
\caption{Time constants $\tau$ for RePLIA at various demodulation frequencies. The central column contains the minimum usable time constant and the right hand column details the actual time constants used which were determined by ensuring a near-Lorentzian passband at the relevant demodulation frequency. Note that this table is not exhaustive and longer/shorter time constants may be used.}
\end{table}

Figure 6a shows an FFT of terminated input data at 10 kHz. These data were extracted from the digital-to-analogue converters (DACs), with the noise level indicating that the DACs themselves are the most significant source of noise. Contrasting this with Figure 3, where we see that without DAC noise, varying demodulation frequency causes a distinct change to noise levels. Figure 7b demonstrates an increase in output signal to noise with increasing demodulation frequency.
\begin{figure}
\centering
\includegraphics[width=0.3\textwidth]
{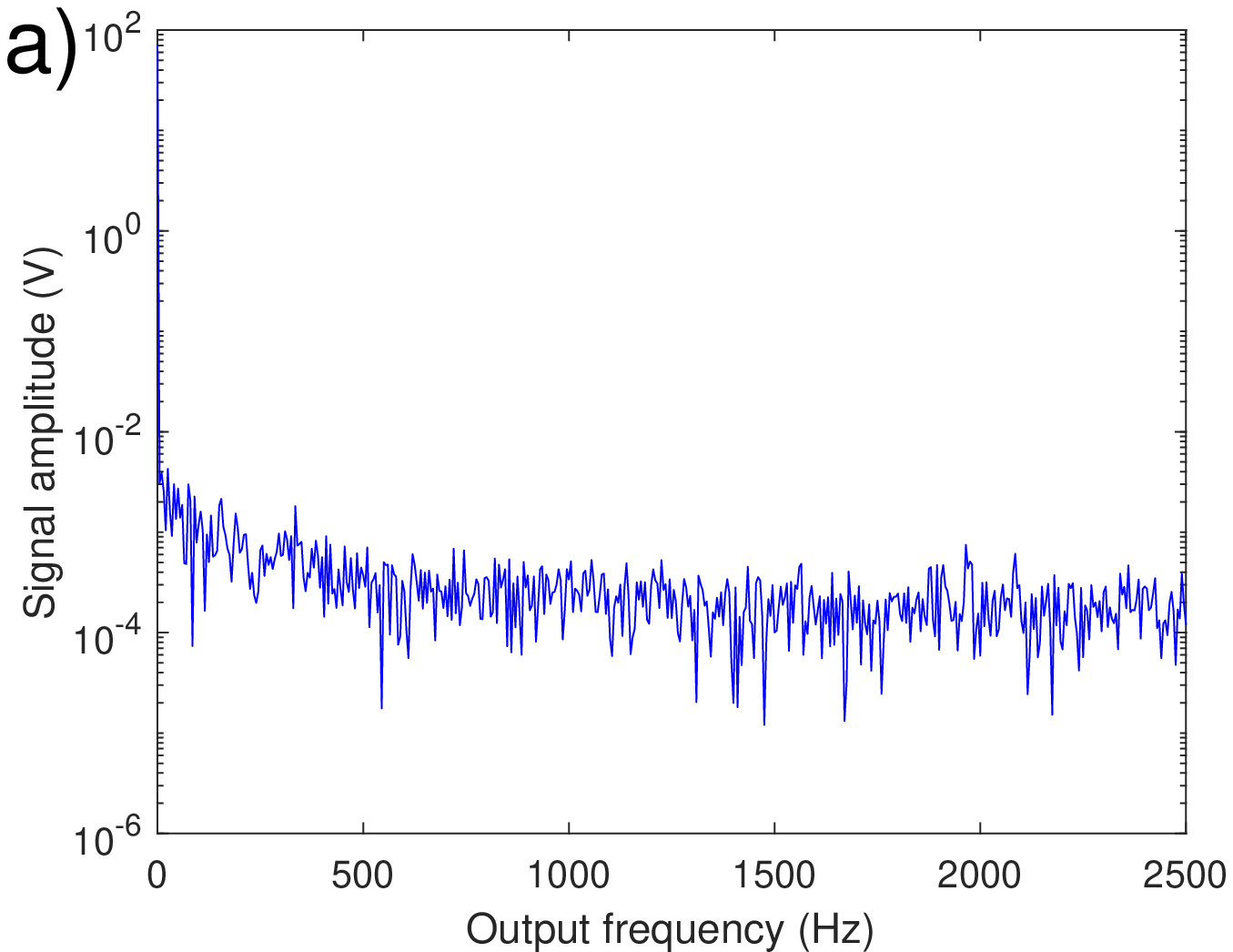}
\includegraphics[width=0.3\textwidth]{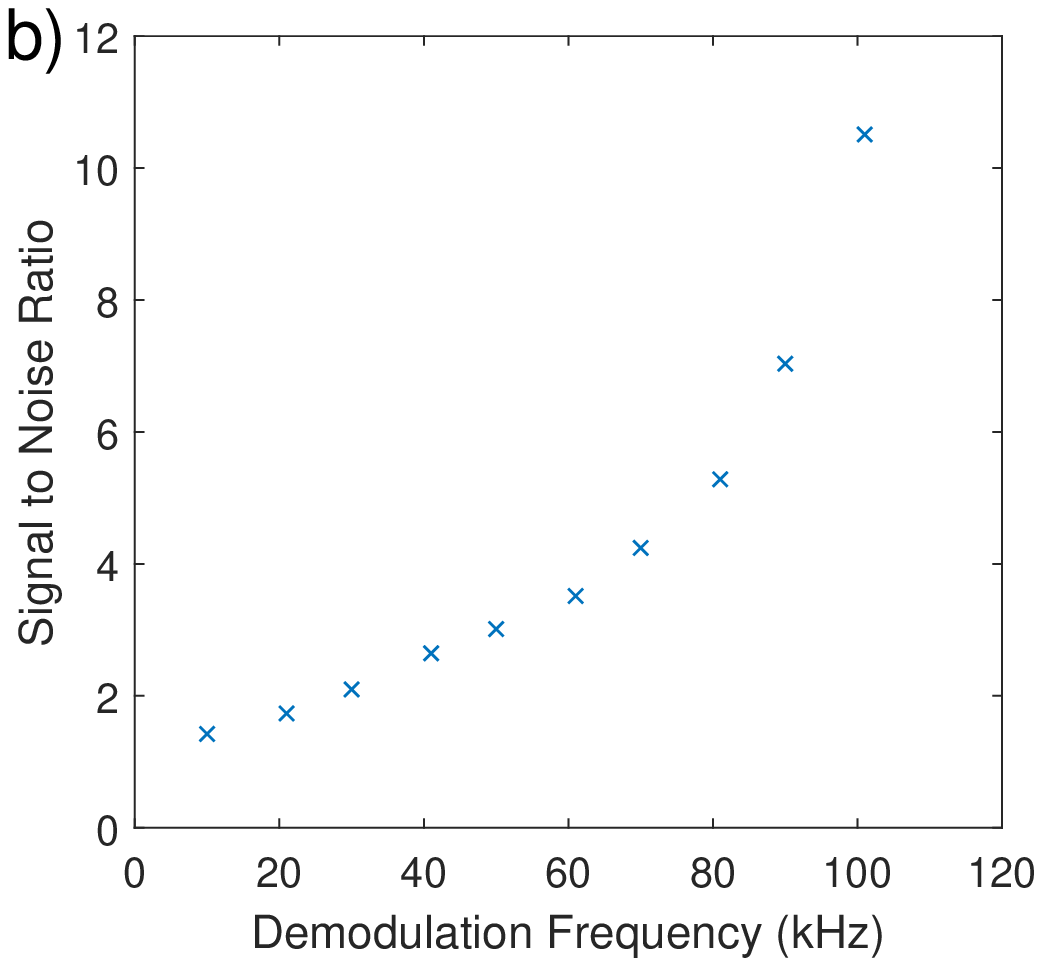}
\caption{a) FFT of 10 kHz demodulation data taken from the RePLIA's DAC outputs, demonstrating significantly greater noise than data taken from Ethernet output. b) Output signal to noise from the DAC outputs improves dramatically with demodulation frequency throughout lower frequency ranges due to increased signal.}
\label{fig:fft10}
\end{figure}

\paragraph*{}
Figure 7a shows the effect on an FFT whilst varying the DAC amplification multiplier, with the extremes in this range shown in figure 7b. There is no large difference in noise level even between the. Though figure 7c shows that an increase in DAC multiplier does improve the signal-to-noise ration, care should be taken not to saturate the STEMlab's $\SI{1}{V}$ maximum output.

\begin{figure}
\centering
\includegraphics[width=0.35\textwidth]{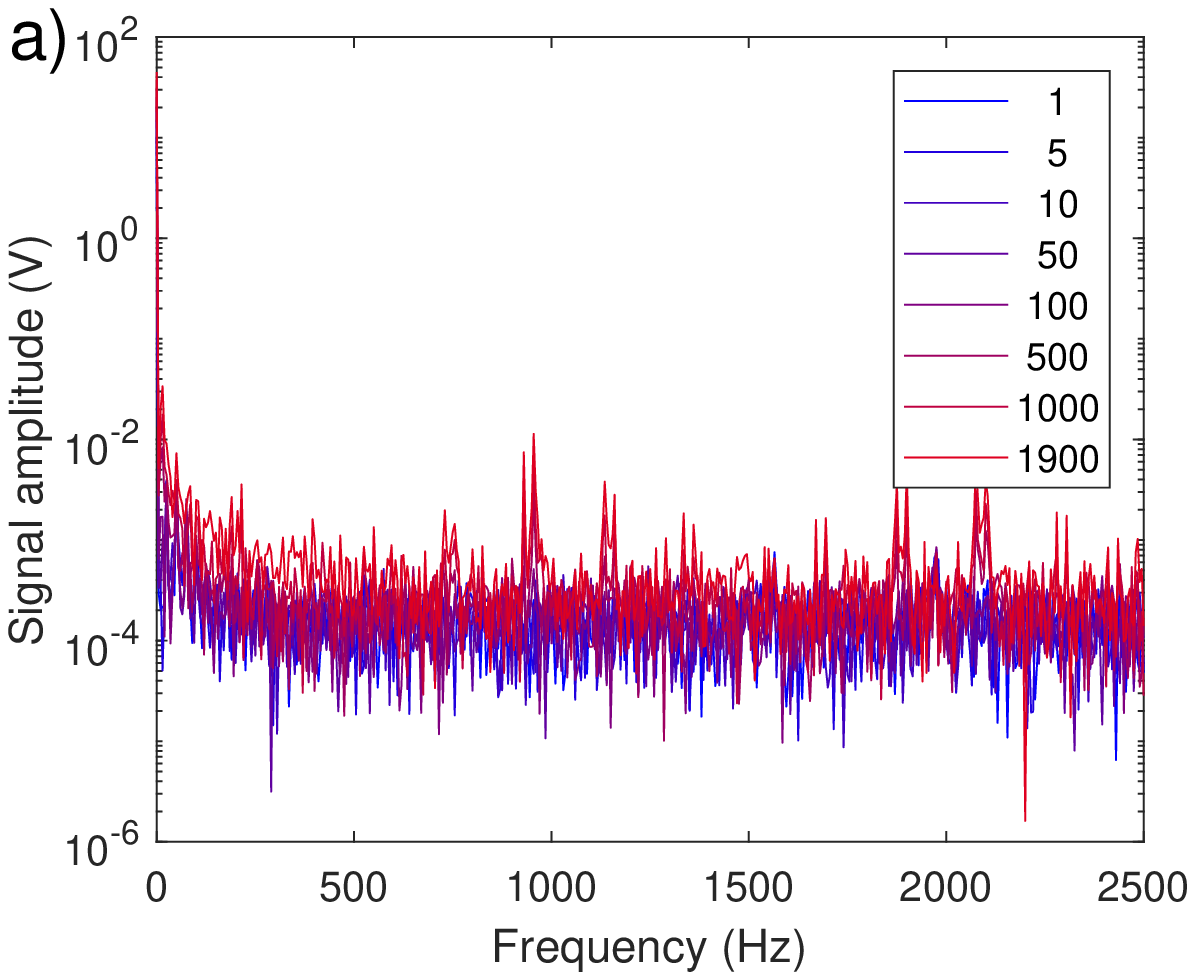}
\includegraphics[width=0.35\textwidth]{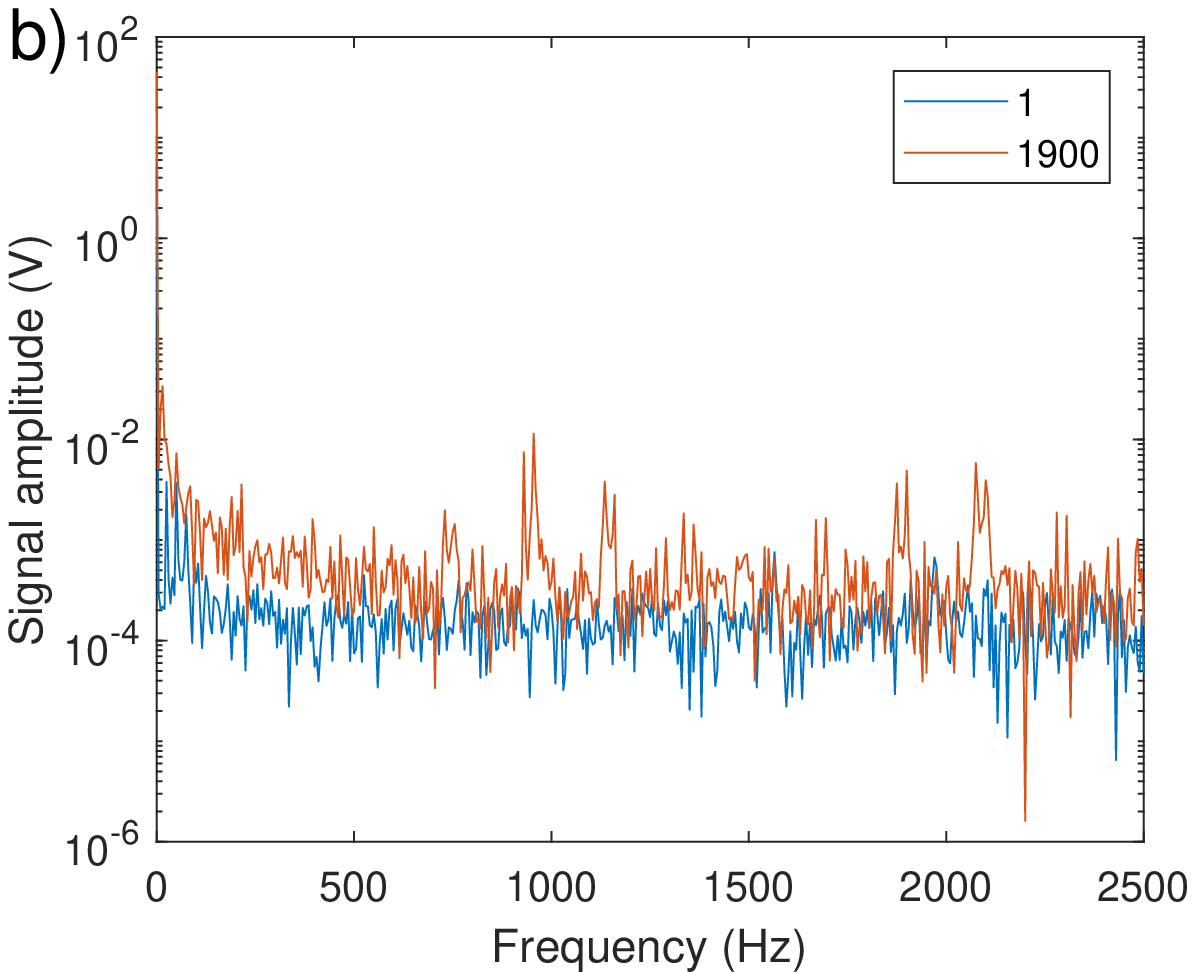}
\includegraphics[width=0.35\textwidth]{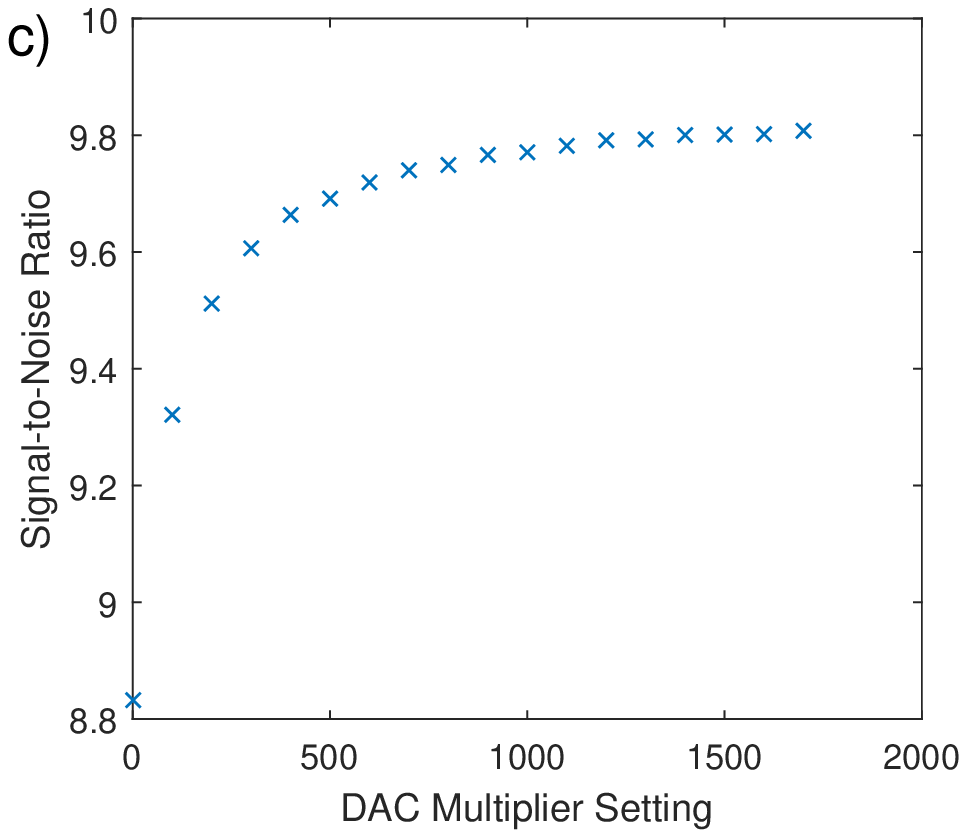}
\caption{FFTs of RePLIA input noise at varying DAC multiplier settings (a) with extremes only plotted for clarity (b). c) The signal to noise ratio at the RePLIA's digital-to-analogue outputs increases with larger DAC multiplier. However, even small signals can lead to saturation of these outputs. Data obtained at $\SI{500}{kHz}$ demodulation frequency and a $\SI{10}{ms}$ time constant.}
\end{figure}

\paragraph*{}
Figure 8 demonstrates the effect of the time constant on the shape of the passband for a given demodulaton frequency, with $\SI{500}{kHz}$ used as an example. Excessively long time constants produce a narrower passband at the expense of line shape; fringes appear at output frequencies close to but greater than the demodulation frequency, leading to the distortion of output signals. Shorter time constants produce a more consistent lineshape at the cost of passband linewidth. Further to this, extremely short time constants (below those detailed in table II) are unable to detect changes on timescales appropriate to the demodulation frequency. In terms of software, the time constant can be set to any value above $\SI{9e-6}{s}$. However, appropriate setting of the time constant is necessary to ensure the correct representation of output signals, and an understanding of the physical nature of systems being measured is also necessary for accurate results.
\begin{figure}
\centering
\includegraphics[width=0.45\textwidth]{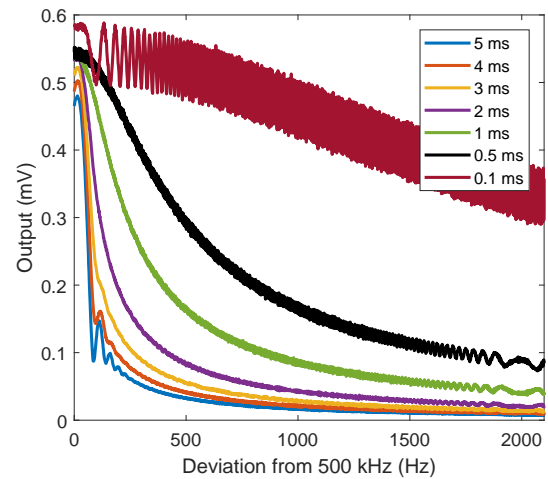}
\caption{The effect of the time constant setting on the resulting passband. An $\SI{80}{kHz}$ sweep over 10 seconds through the demodulation frequency of $\SI{500}{kHz}$ at varying time constants results in distortion of the passband. Longer time constants result in fringes in the unwanted frequency regions and shorter time constants result in excessive acceptance of unwanted frequencies. For 500 kHz, a time constant of $\SI{1}{ms}$ was chosen as a suitable compromise. The small section of the sweep output shown above shows only the part of the passband affected by the time constant.}
\end{figure}
\paragraph*{}
Figure 9 shows FFTs of low demodulation frequency data at varying collection times. Certain frequency elements are evident throughout, which are not evident at higher demodulation frequencies. These components are visible in data extracted both via the DACs and via the STEMlab's Ethernet port. 
\begin{figure}
\centering
\includegraphics[width=0.3\textwidth]
{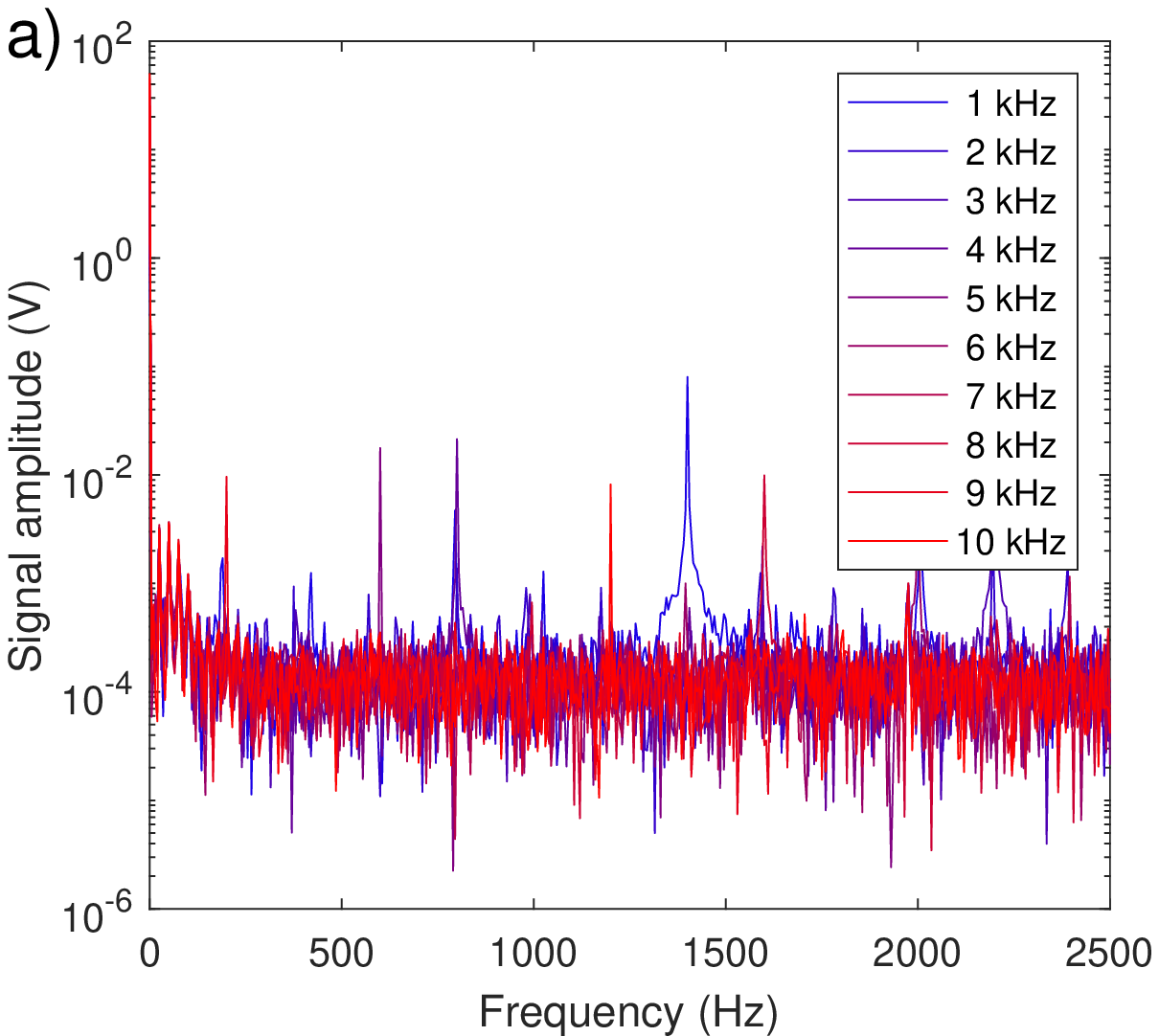}
\includegraphics[width=0.3\textwidth]
{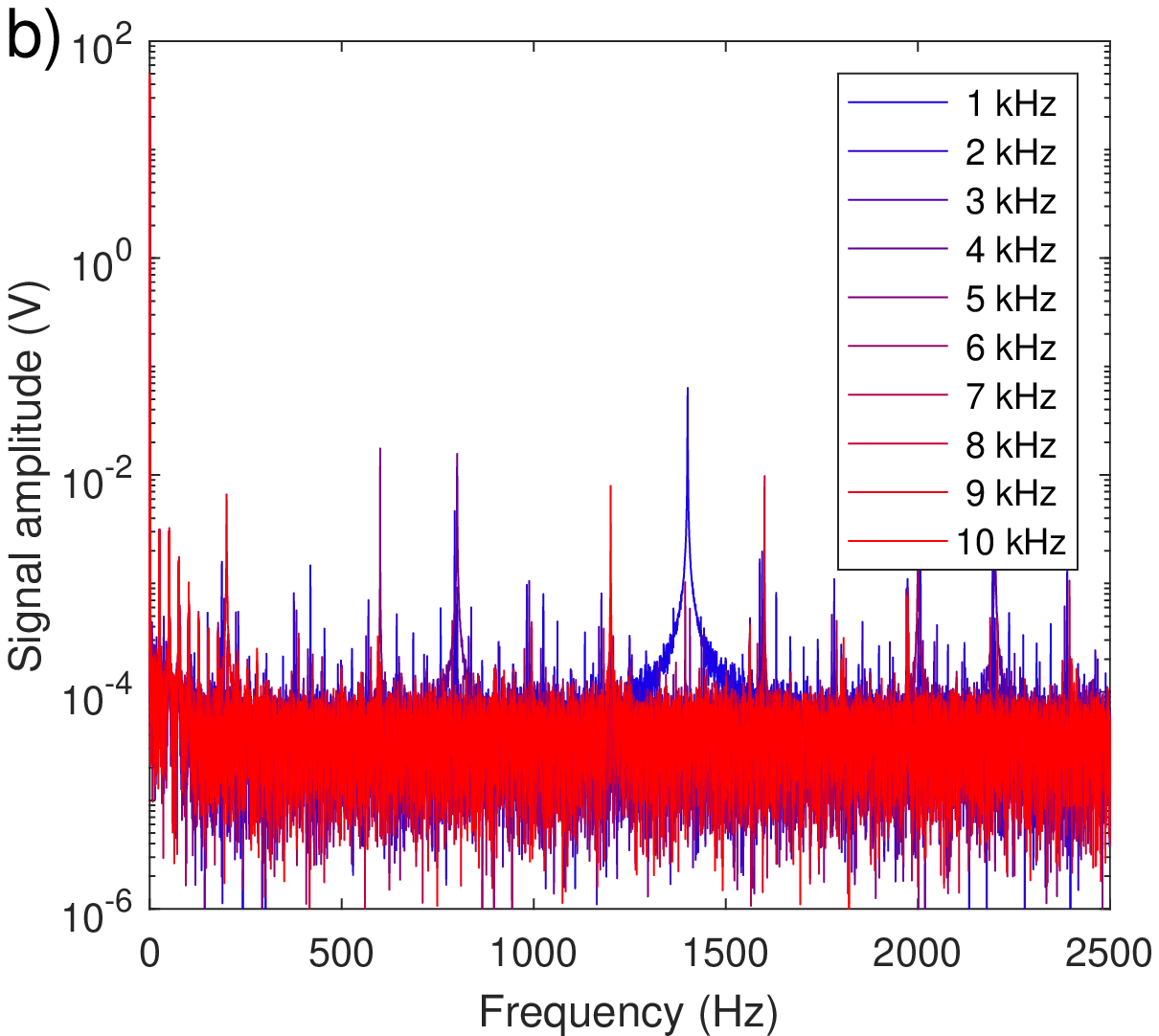}
\includegraphics[width=0.3\textwidth]
{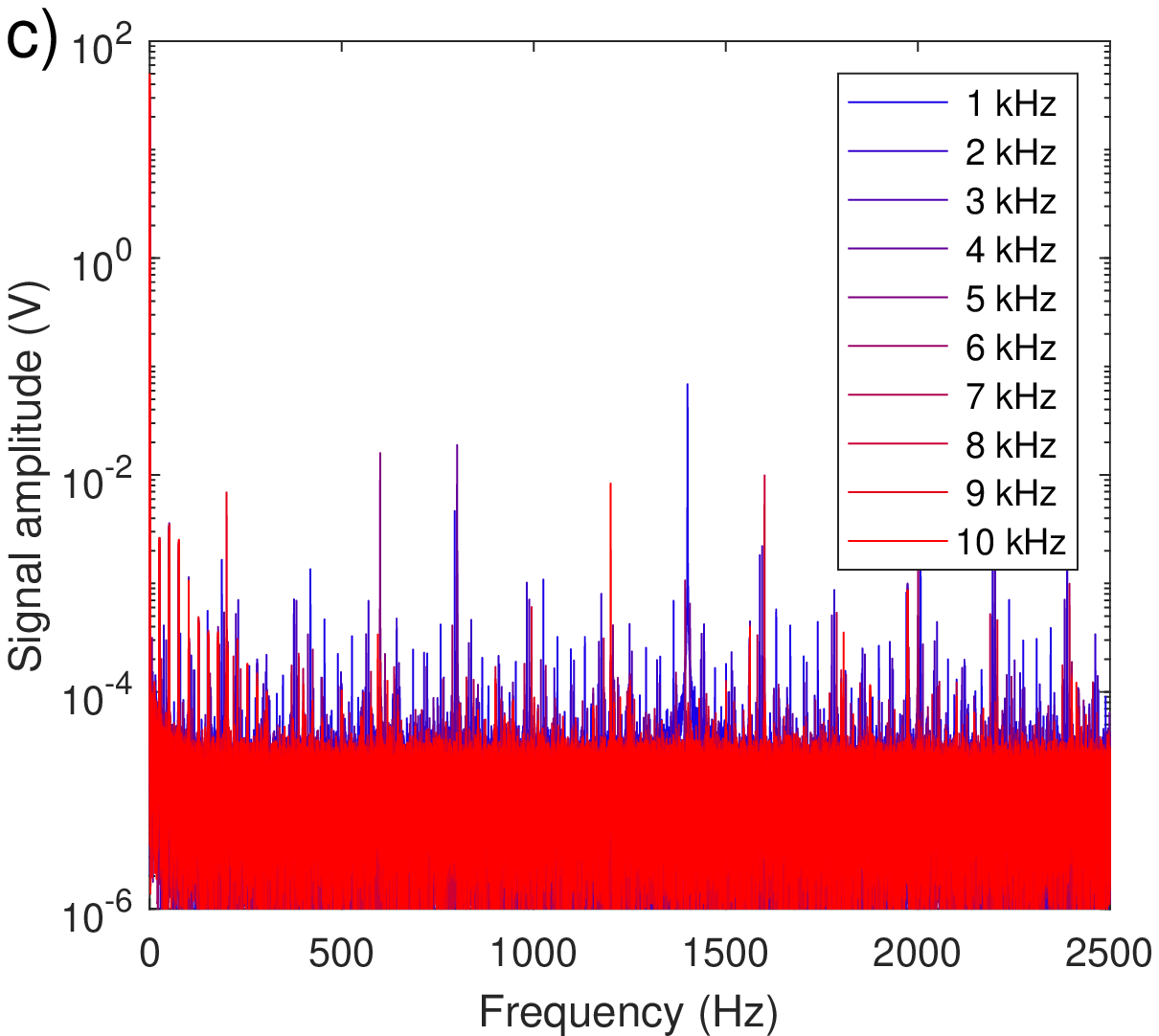}
\caption{Low Frequency (1-10 kHz) FFTs of RePLIA input noise from a) 1 second, b) 10 seconds and c) 100 seconds of DAC data. Frequency-specific spikes are common at low demodulation frequency. }
\end{figure}

Figure 10 shows the effect of time constant on noise. Though noise decreases with longer time constants, signal distortions as seen in figure 8 limit the choice of time constant used.

\begin{figure}
    \centering

\includegraphics[width=0.7\linewidth]{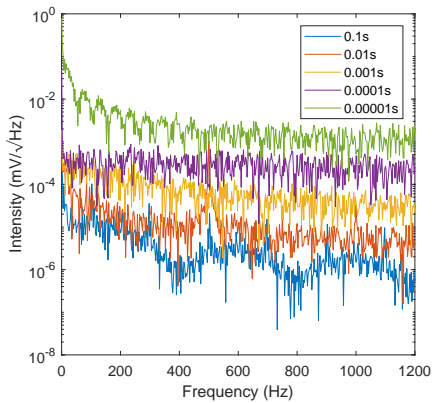}
    \caption{ Note that the 0.00001s trace is below the minimum time constant as detailed in table II.}
    \label{fig:my_label}
\end{figure}
Figure 11 shows the measured $\SI{10}{MHz}$ passband of both the RePLIA and HF2LI along with a Lorentzian fit of the data, demonstrating a rough conformation with such a line shape. This data gives a full-width half-maximum of $\SI{2.6}{kHz}$ for the RePLIA and $\SI{2}{kHz}$ for the HF2LI. Figures 8 and 9 both contain artifacts arising from the single order filter. Increasing the order of the filter may reduce the prominence of these features, although experiments with the HF2LI suggest that such benefits may be marginal.
\begin{figure}
    \centering
    \includegraphics[width=0.6\linewidth]{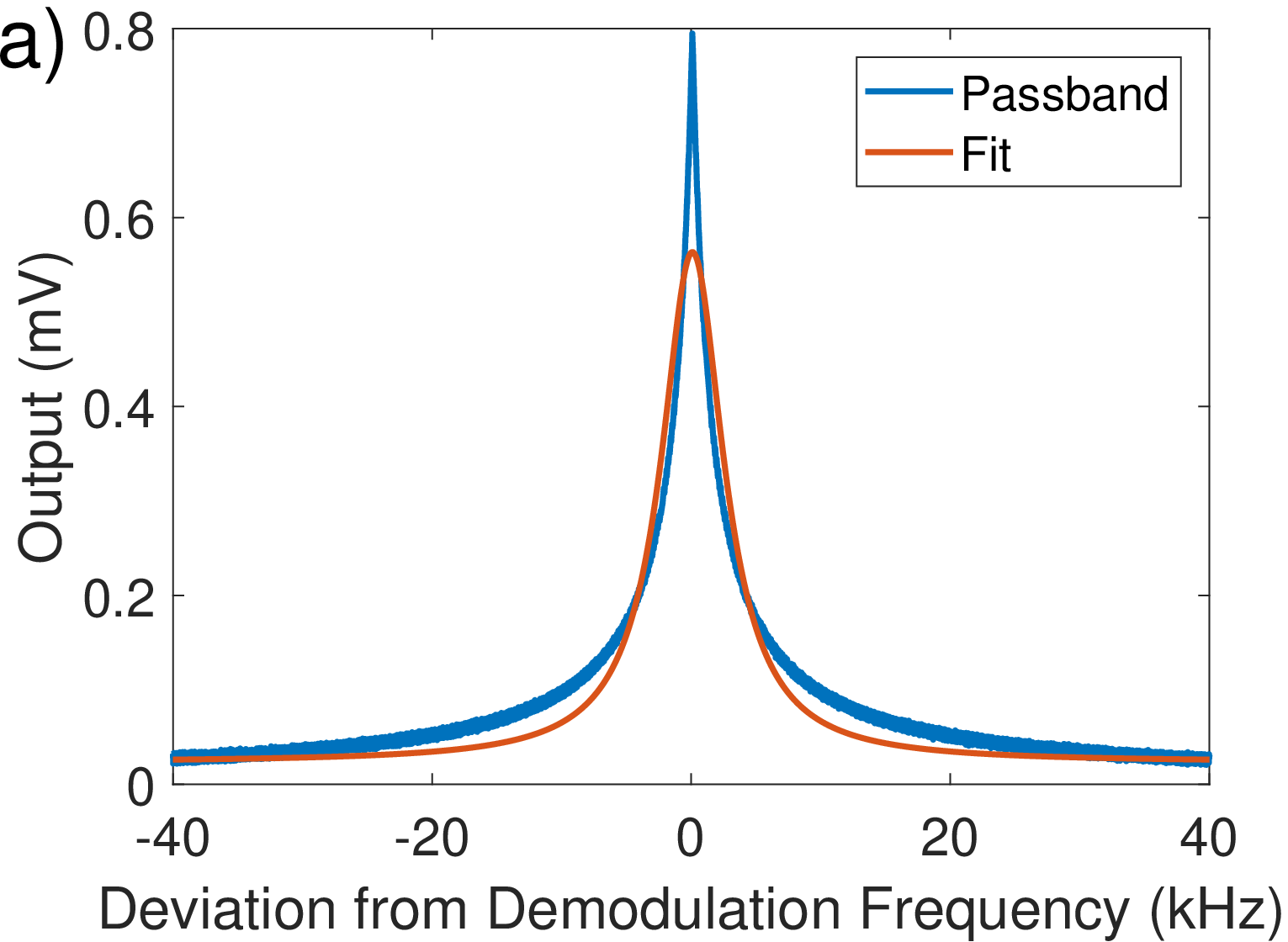}
    \includegraphics[width=0.6\linewidth]{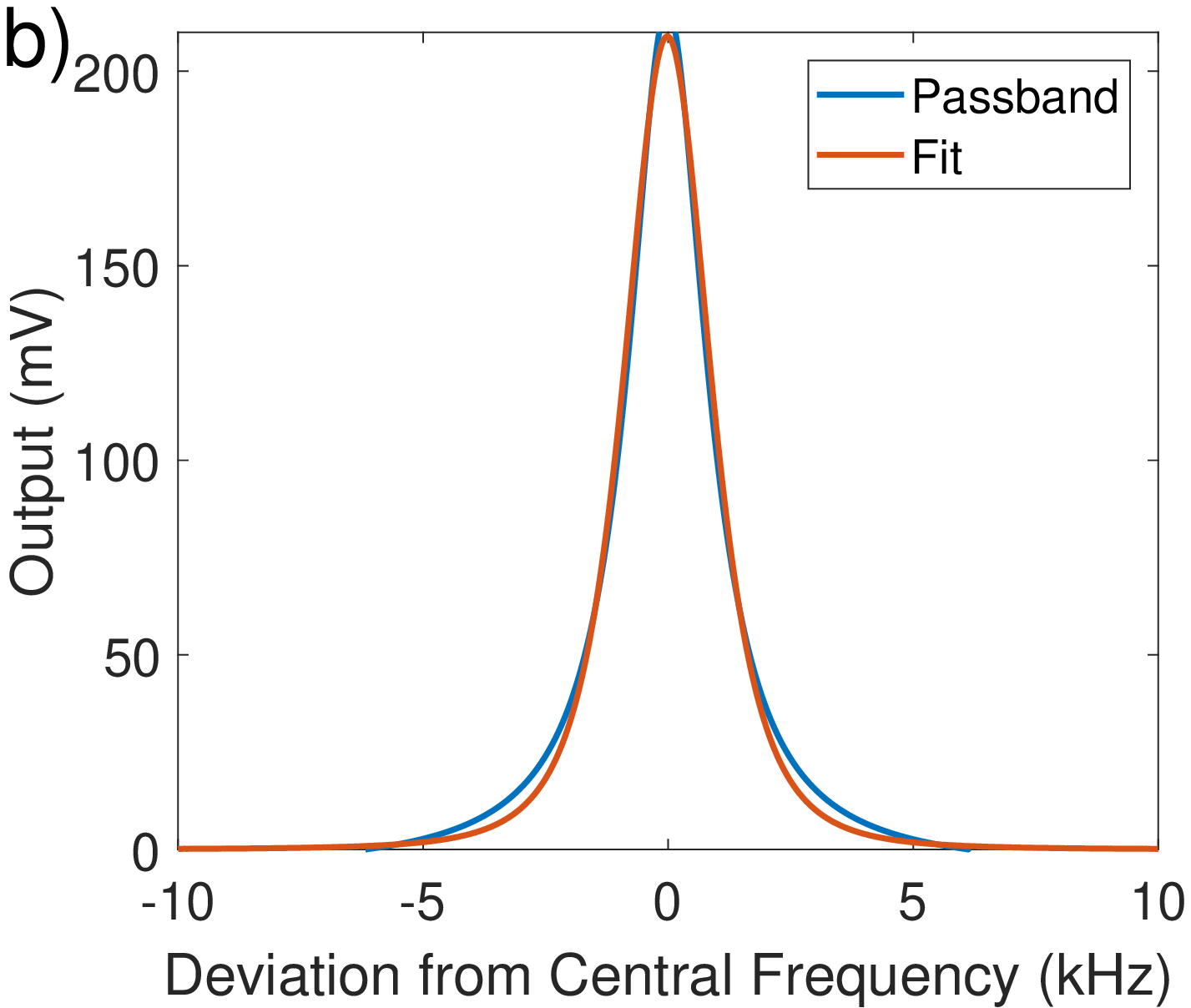}
    \caption{a) RePLIA and b) HF2LI $\SI{10}{MHz}$ passbands with Lorentzian fits with $\SI{2.6}{kHz}$ and $\SI{2}{kHz}$ linewidths respectively. Note that the x-axes cover different ranges for these data.}
\end{figure}

Higher noise at lower frequencies, as seen in figure 3, is consistent with flicker noise of the input stage, as demonstrated in figure 12, in that it is of the form of $1/f$.

\section{Hardware \& Software Notes}
Largely, the RePLIA is limited by factors introduced by the STEMlab's hardware and software. The maximum sample rate of the demodulated signal corresponds to the output sample rate of the STEMlab ($\SI{125}{Ms/s}$) when using the DAC outputs. However, this sampling rate is limited instead by the network connection when when extracting data via the Ethernet adapter. 

Mathematical operations such as $A_{out}=A_{in}-B_{in}$ have not been implemented at the current time, although this is entirely plausible by adding such functionality within the FPGA code.\\\\
Power consumption for the RePLIA is $\SI{4}{W}$ when idle and $\SI{<6}{W}$ during lock in operation.\\
Cross-talk between input channels results in a signal 0.3\% of the input signal amplitude reflected in the secondary input.
Output voltage range is limited by the board to $\SI{\pm 1}{V}$.

\begin{figure}
    \centering
    \includegraphics[width = 0.4\textwidth]{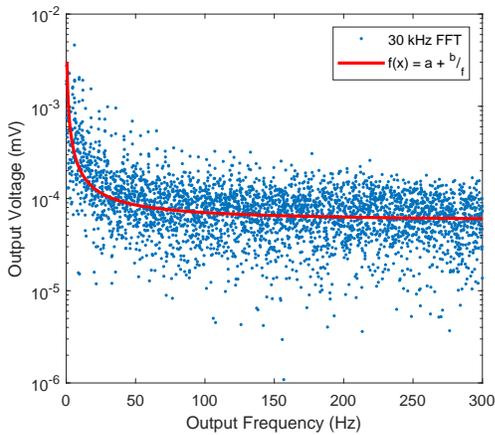}
    \caption{FFT of signal demodulated at 30 kHz, with a fit of $f(x) = a+b\cdot \nicefrac{1}{f}$ where $a = 5.6\cdot10^{-5}, b = 0.001$, consistent with flicker noise of the board.}
    \label{fig:my_label}
\end{figure}
Figure 13 represents the same data as figure 2 in the main text, but as plain FFTs of noise data as opposed to noise spectral density.
\begin{figure}
    \centering
    \includegraphics[width = 0.4\textwidth]{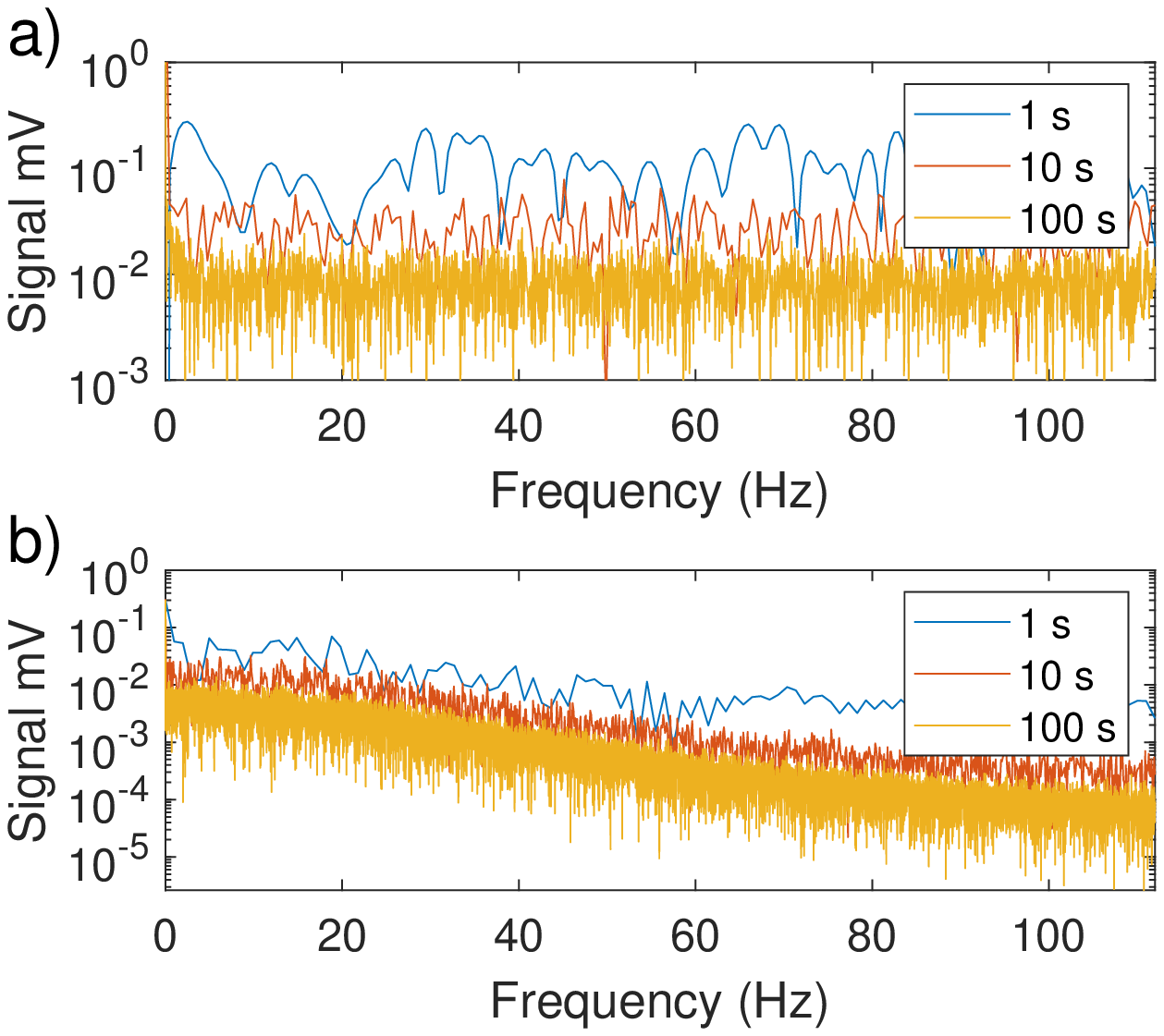}
    \includegraphics[width = 0.4\textwidth]{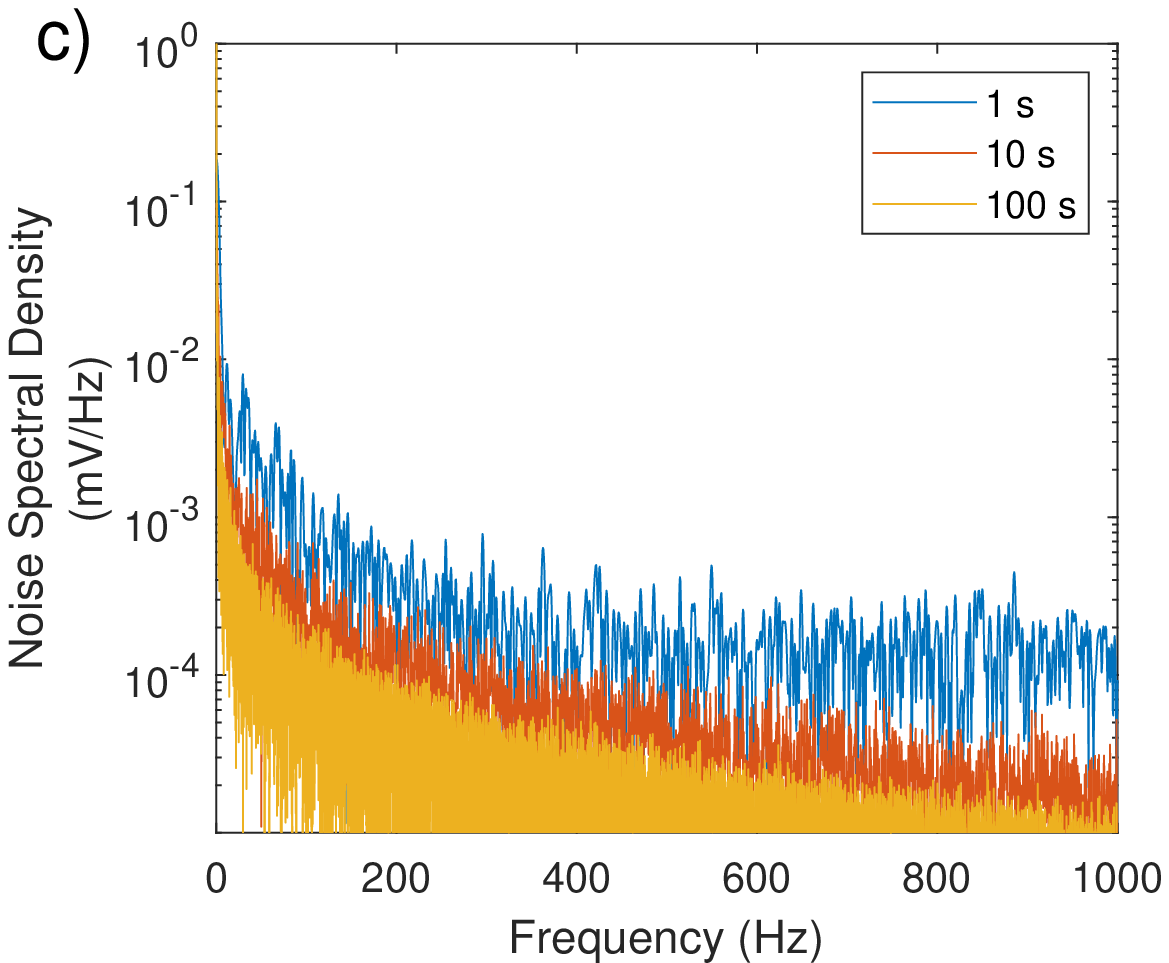}
    \caption{FFTs of noise data from a) RePLIA at $\SI{500}{kHz}$ demodulation with a $\SI{1}{ms}$ time constant (with $\SI{1}{s}$ data zero padded) and b) HF2LI at $\SI{1}{MHz}$ demodulation and $\SI{700}{\micro s}$ time constant, at 1 second (blue), 10 seconds (red) and 100 seconds (yellow) collection time (output data is R for both channels). c) Noise spectral density of the RePLIA at the above input parameters, with a wider output frequency spectrum than in figure \ref{fig:combinednoise}.}
    \label{fig:rpwide}
\end{figure}
\\\\
The RePLIA has been used for magnetometry using an ensemble of nitrogen vacancy centers in diamond. We intend to apply this technique to magnetocardiography, where it would be preferable to use approximately two hundred sensors, requiring one hundred dual channel lock-in amplifiers\cite{morleydale}.
\end{appendix}

\bibliographystyle{unsrt}

\end{document}